\documentclass[prl,aps,showpacs,superscriptaddress,twocolumn,floatfix,
nofootinbib]{revtex4}

\usepackage{amsmath,amsfonts,amssymb}
\usepackage{graphicx}

\def\3{2.8in}    %used for figure widths
\def\2{2.5in}
\def\4{3.0in}

\def \beq {\begin{equation}}
\def \eeq {\end{equation}}

\begin{document}

\title{Development and operation of research-scale III-V nanowire growth reactors}
\author{M. D. Schroer}\altaffiliation{These authors contributed
  equally.}\affiliation{Department of Physics,
  Princeton University, Princeton, New Jersey 08544}
\author{S. Y. Xu}\altaffiliation{These authors contributed
  equally.}\affiliation{Department of Physics,
  Princeton University, Princeton, New Jersey 08544}
\author{A. M. Bergman}\affiliation {Department of Physics,
  Princeton University, Princeton, New Jersey 08544}
\author{J. R. Petta}\altaffiliation{Author to whom correspondence should be addressed; electronic mail: petta@princeton.edu.}\affiliation {Department of Physics, Princeton University, Princeton, New Jersey 08544}

\begin{abstract}
III-V nanowires are useful platforms for studying the electronic and mechanical properties of materials at the nanometer scale. However, the costs associated with commercial nanowire growth reactors are prohibitive for most research groups. We developed hot-wall and cold-wall metal organic vapor phase epitaxy reactors for the growth of InAs nanowires, which both use the same gas handling system. The hot-wall reactor is based on an inexpensive quartz tube furnace and yields InAs nanowires for a narrow range of operating conditions. Improvement of crystal quality and an increase in growth run to growth run reproducibility are obtained using a homebuilt UHV cold-wall reactor with a base pressure of 2 $\times$ 10$^{-9}$ Torr. A load lock on the UHV reactor prevents the growth chamber from being exposed to atmospheric conditions during sample transfers. Nanowires grown in the cold-wall system have a low defect density, as determined using transmission electron microscopy, and exhibit field effect gating with mobilities approaching 16,000 cm$^2$/(V s).
\end{abstract}

\pacs{61.46.Km,73.21.Hb,81.07.Gf}
\maketitle

\section{I. Introduction}
Nanowires - one-dimensional structures with diameters in the nanometer range and lengths of many microns - are ideal systems for determining how the mechanical, electrical and optical properties of materials are impacted by reduced dimensionality and finite-size effects.\cite{Alivisatos_Science_1996,Yakobson_AmScientist_1997} Semiconductor nanowires are of particular interest due to the low effective mass of the charge carriers, which results in Fermi wavelengths on the order of tens of nanometers. In addition, the wires have served as primary components in a wide variety of devices. The one-dimensional nature of nanowires can naturally be applied to make mechanical resonators, where high quality factors, high resonant frequencies, and large scale assembly of devices have been demonstrated.\cite{Husain_APL_2003,Tanner_APL_2007,Li_NatNano_2008} High performance field-effect transistors have been fabricated using Ge/Si nanowires, which exhibit scaled transconductances and intrinsic delays several times better than state of the art Si metal oxide semiconductor field effect transistors.\cite{Xiang_Nature_2006} Boron-doped silicon nanowires have been used to create chemical and biological sensors with sensitivities reaching down to the picomolar concentration range.\cite{Cui_Science_2001,Patolsky_Science_2006} Scaling down in size, few electron quantum dots have been fabricated by locally gating InAs nanowires.\cite{Shorubalko_Nanotech_2007,Fuhrer_NanoLett_2007} In terms of optical devices, nanowire lasers have been built from III-V multi-quantum-well nanowire heterostructures and ZnO nanowires.\cite{Huang_Science_2001,Qian_NatMatl_2008} The ability to grow nanowires of different materials with controlled structure, size and morphology opens up a wide range of fundamental condensed matter physics experiments, as well as the promise of numerous nanoelectronic and nanophotonic applications.\cite{Duan_Nature_2001,Samuelson_PhysE_2004}

Relatively expensive commercial systems are typically used for III-V nanowire growth.\cite{Pfund_Chimia_2006,Dick_JCrystG_2007,Shtrikman_NanoLett_2009} For example, the Aixtron AIX 200 system costs nearly \$1,300,000.\cite{Su_APL_2005,Paiano_JCrystG_2007,Jeppsson_JCrystG_2008} Unfortunately, the cost of commercial reactors presents a high barrier of entry into semiconductor nanowire research for many university-scale research programs. Commercial systems are typically optimized for two-dimensional growth of uniform epilayers over the entire diameter of a wafer, placing very stringent requirements on temperature control and gas flow homogeneity. Given that even a small growth substrate can produce an enormous amount of nanowires (typical densities can be well over 1/$\mu$m$^2$) the design tolerances may be greatly relaxed for a system designed specifically for nanowire growth.

In this article we describe the construction and operation of hot-wall and cold-wall metal organic vapor phase epitaxy (MOVPE) reactors that have been designed to fulfill the need for a relatively simple and low cost system for producing high quality semiconductor nanowires. The article is divided into four main sections. In Sec.\ II we briefly review the vapor-liquid-solid (VLS) nanowire growth process. The construction and operation of the gas handling system, which can be used with either the hot-wall or cold-wall reactor, is described in Sec.\ III. Operation of the hot-wall reactor is then detailed in Sec.\ IV, with emphasis on specific parameters that are important for the growth of InAs nanowires. In Sec.\ V we describe the construction and operation of the UHV cold-wall reactor. Low defect growth of $\sim$60 nm diameter InAs wires with lengths exceeding 5 $\mu$m is demonstrated with this system. Field effect transistors fabricated from these InAs nanowires have field effect mobilities approaching 16,000 cm$^2$/(V s), among the highest values reported in the literature for InAs nanowires.\cite{Jiang_NanoLett_2007,Ford_NanoLett_2009}

\section{II. VLS GROWTH PROCESS}
The VLS growth process, illustrated in Fig.\ 1, was discovered by Wagner and Ellis in 1964 and has been used to grow nanowires made from a wide range of materials.\cite{Wagner_APL_1964,Rao_ProgSSChem_2003,Fan_Small_2006,Dick_JCrystG_2007,Paiano_JCrystG_2007} In standard VLS growth, a substrate is coated with metal nanoparticles, typically gold, which serve as nucleation sites for nanowire growth. The substrate is heated to a high temperature in the presence of III-V source material, which then alloys with the catalyst particles. Under the proper conditions, the liquid catalyst particle becomes supersaturated with source material, and the nanowire material crystallizes atomic layer by atomic layer, lifting the gold particle off of the surface of the growth substrate as the wire grows at the liquid-solid interface. There are many modes of source material delivery, the simplest of which is most likely the solid source vapor transport process.\cite{Huang_AdvMatl_2001,Kuo_JPhysChemB_2006} Here the solid source material is placed in a quartz boat in a hot region of the tube furnace. At high temperature the material vaporizes and then alloys with the catalyst particles on the growth chip, which is held in a lower temperature region of the reactor. Nanowires made out of InAs, ZnO, and many other materials have been grown using this process.\cite{Huang_AdvMatl_2001,Park_APL_2005,Kuo_JPhysChemB_2006,Yan_JPhysChemC_2009} A disadvantage of solid source physical vapor transport, however, is that it is difficult to precisely control the nanowire composition during the growth process. Nevertheless, some groups have developed clever techniques for growing core-shell nanowires using this method.\cite{Jiang_NanoLett_2007}

\begin{figure}[t]
%\centering
\includegraphics[width=8.6cm]{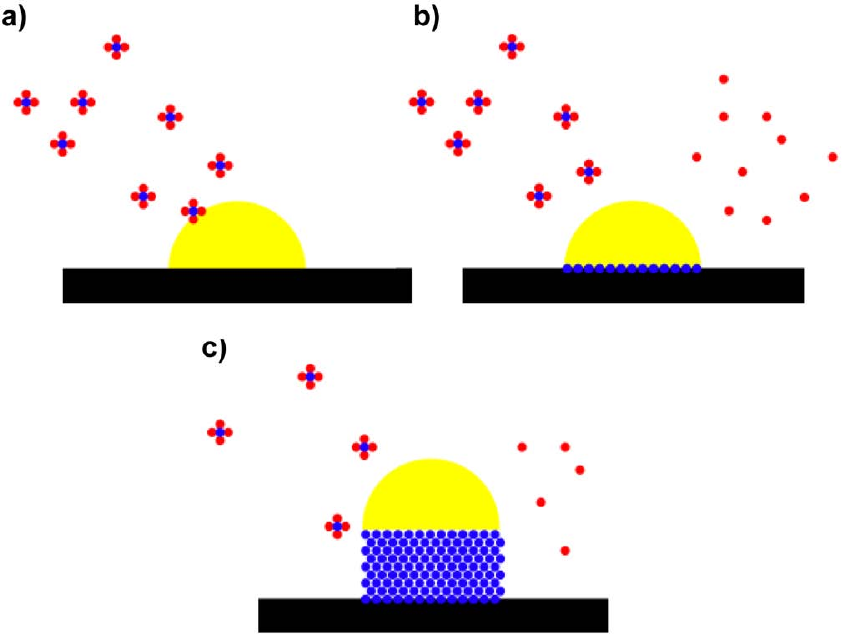}
\caption{The VLS growth process. (a) A gold catalyst particle located on an InAs $<$111$>$B surface is heated in the presence of metal organic precursor. (b) The metal organic precursor preferentially decomposes at the gold catalyst particle. The gold particle becomes supersaturated with In, and nanowire growth proceeds atomic layer by atomic layer. (c) The gold catalyst particle rises as the wire grows. Decomposition of the precursor on the sidewall of the nanowire leads to tapered growth. In most instances, the degree of tapering can be minimized by tuning the growth temperature and precursor partial pressures.}
\end{figure}

\begin{figure}[t]
%\centering
\includegraphics[width=8.6cm]{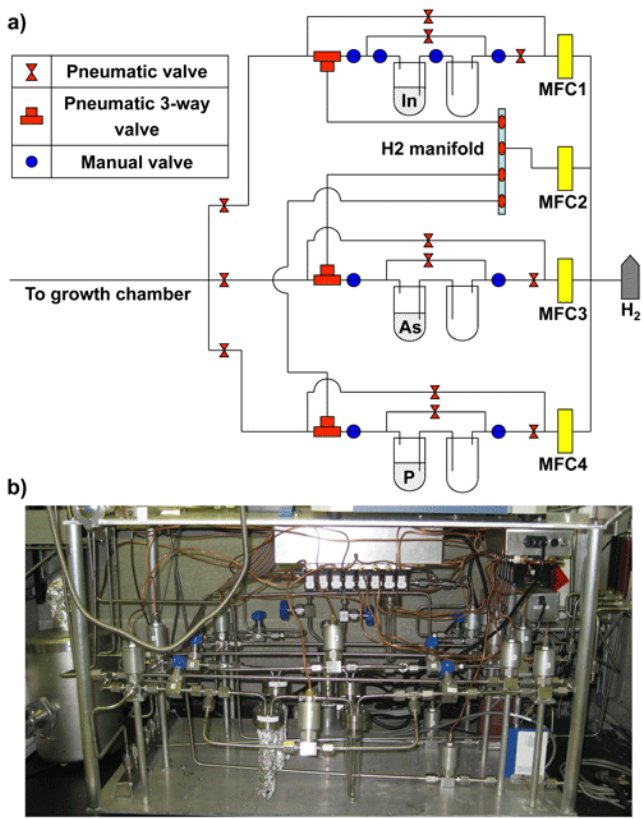}
\caption{Gas handling system used with the hot-wall and cold-wall reactor. (a) Hydrogen carrier gas flow through the three precursor banks is controlled by MFC1, MFC3, and MFC4. MFC2 is designed for a higher carrier gas flow rate and feeds the hydrogen manifold. Hydrogen from the manifold is directed to three-way valves at the end of each manifold and is used to dilute the precursor concentrations. The lines exiting from each precursor bank are then recombined and sent to the growth chamber. (b) Image of the gas handling system, which is designed to fit in a laboratory fume hood.}
\end{figure}
It is desirable to be able to adjust the delivery rate of source materials independently of temperature and pressure in order to expand the range of available growth conditions and provide control of the growth kinetics. Molecular beam epitaxy (MBE), chemical beam epitaxy (CBE), and laser ablation have been adapted for use in the VLS growth process. VLS growth using MBE and CBE has resulted in wires of high crystalline quality and has even been extended to grow one-dimensional heterostructures with clean interfaces.\cite{Bjork_NanoLett_2002,Ohlsson_PhysE_2002} However, most commercial MBE and CBE systems involve large UHV chambers that are very expensive to set up and maintain, placing them out of range of most research budgets. On the other hand, chemical vapor deposition (CVD) systems are operated at pressures in the range of 1-100 Torr, making the operation simpler and more affordable.\cite{Stringfellow,Dick_JCrystG_2007} Nevertheless, typical CVD growth usually requires hazardous gas phase precursors including silane, phosphine, and arsine, all of which are toxic at extremely low concentrations. MOVPE, also referred to as metal organic chemical vapor deposition, uses liquid-phase chemicals that tend to form less toxic metal oxides upon reaction with air. For reasons of laboratory safety, and due to the fact that the number of commercially available metal organic precursors is rapidly expanding, we focus on the development of MOVPE nanowire growth reactors.

\section{III. GAS HANDLING SYSTEM}
The hot-wall and cold-wall reactors are supplied with metal organic precursors using the gas handling system shown in Fig. 2(a).  The gas handling system was designed to be simple, while maintaining flexibility and keeping costs down. All the components for this system may be purchased for around \$28,000, depending on the choice of mass flow controllers (MFCs) and their control interface. As all welding and assembly were performed in-house there were no additional costs, making this a very affordable system.

An image of the as-built gas handling system is also shown for comparison in Fig.\ 2(b). Gas flows in this diagram from right to left. All tubing is electropolished stainless steel with 1/4 in.\ VCR fittings. High purity (99.9999\%) hydrogen gas at a pressure of 40 psi is sent to a four-way splitter. Each output from the splitter feeds a MFC. Three banks of bubblers are designed to supply the precursors: trimethylindium (TMI), triethylarsine (TEAs) and di-tert-butylphosphine. MFC1, MFC3, and MFC4 are MKS model 1179 mass flow controllers configured to flow a low rate [$<$50 SCCM (SCCM denotes cubic centimeter per minute at STP)] of hydrogen gas through each bank of bubblers. Instead of using a single bubbler for each bank, we place two bubblers back to back as shown in Fig.\ 2(a). Any accidental pressure imbalance that would normally drive the liquid precursor into the MFC is caught by the secondary bubbler. The manual Swagelok model 6LVV-DPFR4-P-C valves in the gas handling system are used to slowly bring the pressure of the precursors into equilibrium with the growth chamber. Output from MFC2, which has a full scale range of 1000 SCCM, is sent to a manifold that splits the gas into three paths, providing hydrogen gas that is used to dilute the precursor concentration immediately downstream of the bubblers. The connection to the output of each bubbler bank is made using a Swagelok 6LVV-DPA222-P-C three-way pneumatic valve. One last set of Swagelok 6LVV-DPMR4-P-C pneumatic valves separates the diluted output of each bubbler bank from the growth chamber. The pneumatic valves are capable of withstanding a short term bakeout of 150 $^\circ$C. All of the pneumatic valves are actuated using compressed air lines controlled by Humphrey MB3E1 solenoid valves, which are in turn driven by Omron G3M-203P DC5 solid state relays. A 96 channel National Instruments PCI-6509 digital card sources a sufficient amount of current (24 mA/channel) to drive the relays. Nanowire growth is largely automated by using a National Instruments LabView program to control all of the valve sequencing.

The growth pressure for both the hot-wall and cold-wall systems is controlled using downstream feedback. The pressure is measured at the growth chamber using a MKS Instruments model 121 capacitance manometer. This is interfaced with a MKS 146C vacuum gauge controller, which controls a MKS 148J proportional valve located on the exhaust line. The proportional valve has a conductance that works well for growth pressures between 30--760 Torr with the gas flow rates used here. Lower growth pressures could be attainable by substituting a higher conductance valve.  The proportional valve can be bypassed using a Swagelok 6LVV-DPMR4-P-C on/off pneumatic valve for rapid evacuation of the system. The hot-wall system uses a RV8 pump equipped with standard ``ultragrade 19" oil. A larger Edwards RV12 mechanical pump, designed for use with low vapor pressure pump oil (Fomblin 06/6), was connected to the cold-wall reactor. A Nor-Cal FTN-6-1502-NWB 6 l liquid nitrogen cold trap is placed between the Edwards RV pump and the proportional valve in order to minimize liquid nitrogen boiloff. The trap has sufficient capacity to last through an entire growth run.

\section{IV. HOT-WALL GROWTH SYSTEM}
We constructed a simple hot-wall reactor based on a quartz tube furnace. The furnace, made by Lindeberg/Blue M, is configured to hold an $\sim$24 in.\ long 1 in.\ diameter quartz tube and reaches a maximum temperature of 1100 $^\circ$C.  The output from the gas handling system is sent directly to the quartz tube and the pressure in the system is controlled using the downstream feedback system described in Sec.\ III.  The only cost to integrate this reactor with the gas handling system of Sec.\ III is a tube furnace and a few miscellaneous vacuum parts, for a total of price of around \$2000. A typical growth run consists of the following steps:

\begin{figure}[t]
%\centering
\includegraphics[width=8.6cm]{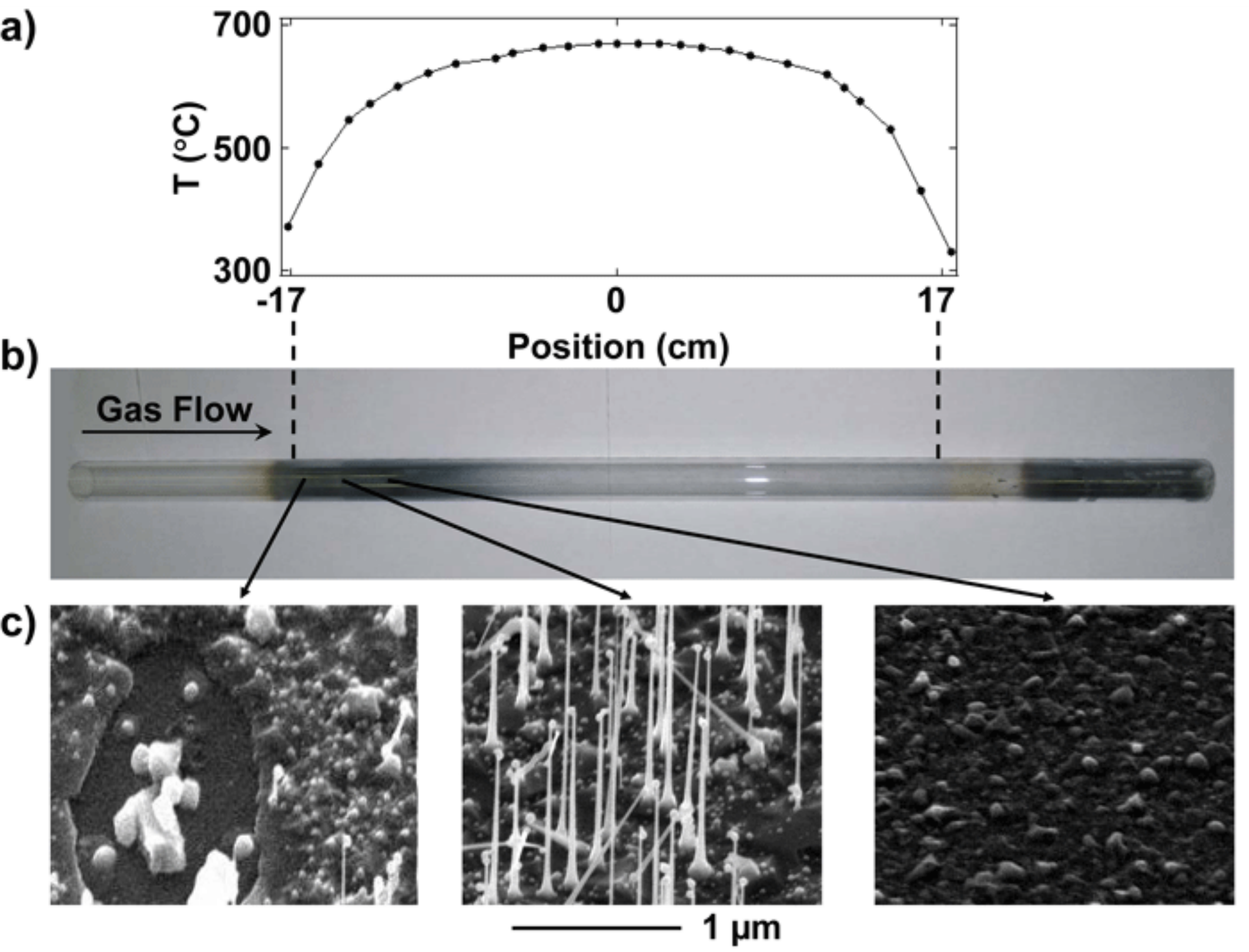}
\caption{Hot-wall reactor. (a) Temperature profile of the hot-wall furnace measured using a thermocouple placed inside of the quartz tube. (b) Image of the quartz tube after several InAs nanowire growth runs. Gas flow in the tube is from left to right. A heavy band of deposition is visible on the left and right ends of the tube. (c) InAs substrates placed near the left end of the tube resulted in InAs nanowire growth with a yield that is highly position dependent due to the steep temperature gradient in this region.}
\end{figure}

\begin{enumerate}
\item Sample preparation-- The InAs $<$111$>$B growth substrate, purchased from Nova Electronic Materials, is etched in 1:10 HCl:H$_2$O solution for 30 s and rinsed in deionized water. A drop of Poly-L-Lysine is then applied to the surface of the chip to enhance adhesion of the gold nanoparticles. After 30 s, this is blown off with dry nitrogen. Finally, Ted Pella 40 nm gold colloid is dropped on the chip, and blown off with dry nitrogen after 10 s. The prepared sample is then immediately loaded into the tube furnace.
\item Pump and purge-- After the chip is loaded, the tube is evacuated to a pressure of 0.5 Torr and then purged with hydrogen to a pressure of 200 Torr. The pump and purge cycle is repeated three times.
\item Precursor load-- The metal organic precursors are carefully brought up to the growth pressure by opening the manual valves on both sides of the bubblers.
\item Postload purge-- The system is purged with hydrogen gas at a flow rate of 600 SCCM for 5 min. The pressure is then stabilized at 50 Torr.
\item Annealing- A 5 min. anneal step is performed before the growth is initiated by ramping the temperature to 600$^\circ$C and flowing hydrogen through the TEAs bubbler at a rate of 10 SCCM (resulting in a partial pressure p$_{TEAs}$=1.1 $\times$ 10$^{-4}$ atm). The anneal desorbs the native oxides on the substrate, melts the gold catalyst particles, and forms a Au-In eutectic alloy.\cite{Hiruma_JAP_1995}
\item Growth-- Nanowire growth is initiated by lowering the furnace temperature to the growth temperature, $\sim$500 $^\circ$C for InAs. Carrier gas flow through the TEAs bubbler remains the same during the temperature sweep. Once a stable growth temperature has been reached, carrier gas is sent through the TMI bubbler at a rate of 2.5 SCCM (partial pressure $p_{TMI}$= 9.1 $\times$ 10$^{-6}$ atm). Typical growth runs last for 30 min.
\item Growth termination-- Carrier gas flow through the TMI bubbler ceases at the end of the growth run. The flow rate through the TEAs bubbler is held steady until the growth system temperature reaches $\sim$400 $^\circ$C. After this the manual valves on the precursors are closed and hydrogen flow is maintained to flush the remaining precursors from the system.
\item Sample removal-- When the system reaches $\sim$80 $^\circ$C, hydrogen flow is terminated. The system is pumped to 0.5 Torr and refilled with hydrogen to a pressure of 760 Torr. The sample is then removed.
\item Placement into standby mode-- The growth chamber is pumped down to 0.5 Torr and left under vacuum.
\end{enumerate}

Initial growth runs were performed with the substrate located at the center of the tube furnace, where the temperature is relatively insensitive to the position of the chip [see Fig.\ 3(a)]. Growth temperatures were varied from 380--500 $^\circ$C for a wide range of precursor concentrations. However, no combination of growth parameters was found to produce InAs nanowire growth. We attribute this to gas phase reactions and depletion of the precursors on the wall of the tube furnace.\cite{Mountziaros_JElecSoc_1991} In contrast, we successfully grew InP wires on chips placed in the center of the tube for a wide range of growth conditions. The sensitivity of InAs nanowire yield on the growth conditions has also been reported in the literature by Dick \textit{et al}.\cite{Dick_AdvFuncMatl_2005,Dick_NanoLett_2005,Dick_JCrystG_2007} In order to compensate for the precursor depletion, the substrate was moved closer to the gas inlet, at a location in the quartz tube that had deposition on the walls [see Fig.\ 3(b)]. While this strategy improves the InAs nanowire growth chemistry, it sacrifices precise control of the growth temperature, since the chip is now located in a region with a large temperature gradient.

Figure 3(c) shows the results of a run with $p_{TMI}$ = 9.1 $\times$ 10$^{-6}$ atm, p$_{TEAs}$ = 1.1 $\times$ 10$^{-4}$ atm, and T$_{growth}$ = 500 $^\circ$C, where the temperature is measured at the center of the quartz tube. Each growth substrate is separated from the next by 6 mm. This run is typical of those performed with growth chips near the inlet of our tube furnace; nearly every run produces nanowires on at least one of the growth chips. Table I quantifies the results from one such successful growth. However, Fig.\ 3(c) also demonstrates the difficulty in optimizing growth using this method; since the growth results are strongly position dependent, the specific morphology of the nanowires produced (degree of tapering, growth rate, and surface roughness) was observed to vary greatly from one run to the next. An additional disadvantage of a tube furnace reactor is that it is not easy to implement a load-lock system. As a result, the entire reactor is exposed to atmosphere before and after each growth run, compromising the purity of the process. While we were occasionally able to produce clean, relatively untapered nanowires with this system, the results demonstrate that there is significant room for improvement.

\begin{table*}
\caption{\label{tab:table1}Nanowire statistics from hot-wall and cold-wall growths.  Each growth was performed for 30 min with the parameters described in the text.  To produce the values below, the length of 50 nanowires and the diameter of 25 nanowires were measured from each run using a SEM.}
	\begin{ruledtabular}
	\begin{tabular}{lcccc}
	
		 & & Standard deviation & & Standard deviation \\
	 	 & Average length & of length & Average diameter & of diameter\\
		 & ($\mu$m)&($\mu$m)&(nm)&(nm)\\
		 \hline
		 Hot-wall system&1.0&0.4&23&7\\
		 Cold-wall system&6.6&0.4&65&5\\

	\end{tabular}
	\end{ruledtabular}
\end{table*}

\section{V. COLD-WALL GROWTH SYSTEM}
By design, a cold-wall reactor can overcome some of the difficulties associated with the hot-wall reactor. In contrast to the hot-wall reactor, only the sample is heated. Therefore pyrolysis and reactions involving the precursors are guaranteed to take place either at the substrate itself or in the stagnant gas layer directly above the substrate. In addition, gas phase reactions and precursor depletion are greatly suppressed, leading to tighter control of the reaction chemistry. In general, we wanted to design a reactor capable of achieving base pressures of less than 1 $\times$ 10$^{-8}$ Torr using standard off-the-shelf components. As a result, the reactor chamber is of an entirely different design compared to the hot-wall system, and represents a much larger investment. Our total cost for the reactor and additional pumps was approximately \$56,000, although this can vary greatly based on various design choices. At the same time, it was important to use the existing gas handling and pressure control system.

\begin{figure}[t]
%\centering
\includegraphics[width=8.6cm]{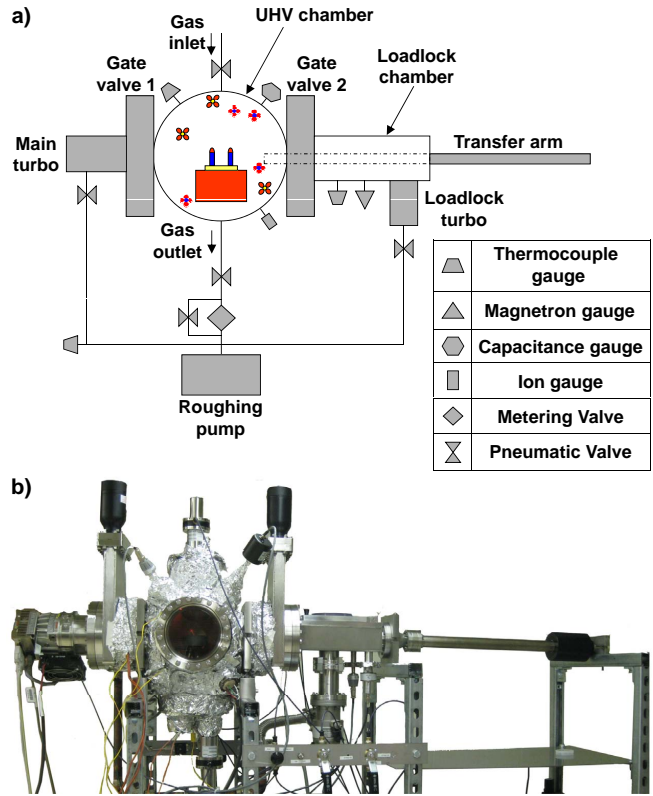}
\caption{Cold-wall UHV reactor. (a) Schematic diagram of the reactor. Nanowire growth takes place in a K.\ J.\ Lesker spherical chamber. The substrate heater is mounted on the bottom flange. The chamber can be isolated from the turbo pump during growth using gate valve 1. Gate valve 2 is opened during sample transfers. A small turbo pump is used to evacuate the load-lock chamber before sample transfers. (b) Image of the assembled cold-wall reactor. The length of the system, measured from the left end of the main chamber turbo pump to the right end of the transfer arm is $\sim$2 m. The chamber is supported on a homemade Unistrut frame. The roughing pump (not shown) is located in a nearby fume hood.}
\end{figure}

Our cold-wall reactor design is shown in Fig.\ 4(a) and is built around a K.\ J.\ Lesker SP1200S 12 in.\ spherical chamber. The spherical chamber has four 8 in.\ conflat ports, two 6 in.\ conflat ports, one 4.5 in.\ conflat port, and four 2.75 in.\ conflat ports, easily accommodating all pressure gauges, valves, and feedthroughs needed to run the system. High vacuum is achieved by pumping on the spherical chamber with a 210 l/s Varian V301 turbo pump, which is separated from the main chamber by a 6 in.\ K.\ J.\ Lesker SG0600PCCF copper bonnet gate valve. A second 6 in.\ gate valve separates the growth chamber from the load-lock chamber. The top of the chamber is capped with a 6--2-3/4 in.\ conflat adapter flange. A K.\ J.\ Lesker SA0150PCCF copper bonnet valve is mounted to the top flange, labeled ``gas inlet" in Fig.\ 4(a) and electropolished stainless steel tubing runs from the copper bonnet valve to the outlet of the gas handling system. An 8 in.\ diameter conflat flange is mounted to the bottom port of the spherical chamber and was machined to accommodate a pumping line, a thermocouple feedthrough, and a high current feedthrough for the sample heater. A MeiVac SU-200-IH 2 in.\ sample heater was mounted to the baseplate using machined stainless steel rods. The pumping line is opened and closed using a K.\ J.\ Lesker SA0150PCCF copper bonnet valve.

Sample transfer is achieved using a load-lock system. We purchased a complete K.\ J.\ Lesker LLC-PLNR4 load-lock chamber, which is evacuated using a 56 l/s Pfeiffer Balzars TPU-060 turbo pump. Load-lock pressure is monitored using Varian IMR-100 magnetron and L9090305 thermocouple gauges. A Transfer Engineering DBLRP-18 magnetically coupled transfer arm is used for x-axis sample translation, while height control is achieved using a ALDEF-10 dynamic end-effector. A custom quartz sample holder was fabricated by GM Associates Inc. and is designed to rest securely on the MeiVac sample heater with two machined slots that accept the transfer arm mounted fork. The top of the quartz sample holder has a rim along its interior that allows a 2 in.\ wafer to be placed directly above the sample heater. In most cases, we placed a 5 $\times$ 5 mm InAs chip in the center of the 2 in.\ silicon wafer.

Pressure in the main chamber is monitored during growth runs using a MKS instruments model 221 capacitance manometer. The growth pressure is adjusted by tuning the pumping rate with a MKS 148J proportional valve, which is controlled with an MKS 146C vacuum gauge controller. The proportional valve can be bypassed for rapid chamber evacuation using a Swagelok 6LVV-DPMR4-P-C pneumatic valve. The roughing pump is connected to the back of the main chamber turbo pump, the back of the load-lock chamber turbo pump, and to the proportional valve. Pneumatic valves allow the back of these turbo pumps to be isolated from the roughing pump during growth runs. Pressure in the roughing pump line is monitored using a Varian L9090306 thermocouple gauge.

The main chamber is kept under high vacuum in standby mode between growth runs. Pressure is then monitored with a Varian UHV-24P conflat flange mounted ion gauge. A Varian XGS600H1M1C1 vacuum controller is used to monitor the load-lock chamber pressure and the main chamber pressure. The Varian and MKS vacuum controllers are interfaced to the control computer using a Quatech ESU2-400 8 port RS232/RS485 module.

As this reactor is designed to be operated by students in an academic setting, safety interlocks were installed whenever possible to prevent accidents. The interlocks were implemented in hardware rather than software to ensure correct operation even in the event of a computer malfunction. Each interlock consists of a set of logic encoded by a series of normally closed or normally open solid state relays that are driven by various setpoint signals. Important setpoint signals include the pressure in various parts of the system, the state of the turbo pump and the status of other valves. For instance, the valve connecting the roughing pump to the chamber may only be opened if the chamber pressure is above 1 mTorr (to prevent oil backstreaming) and if both of the turbo pump foreline valves are closed. The interlocks result in a fairly robust system that is resistant to a low level of operator error.

\begin{figure}[t]
%\centering
\includegraphics[width=8.6cm]{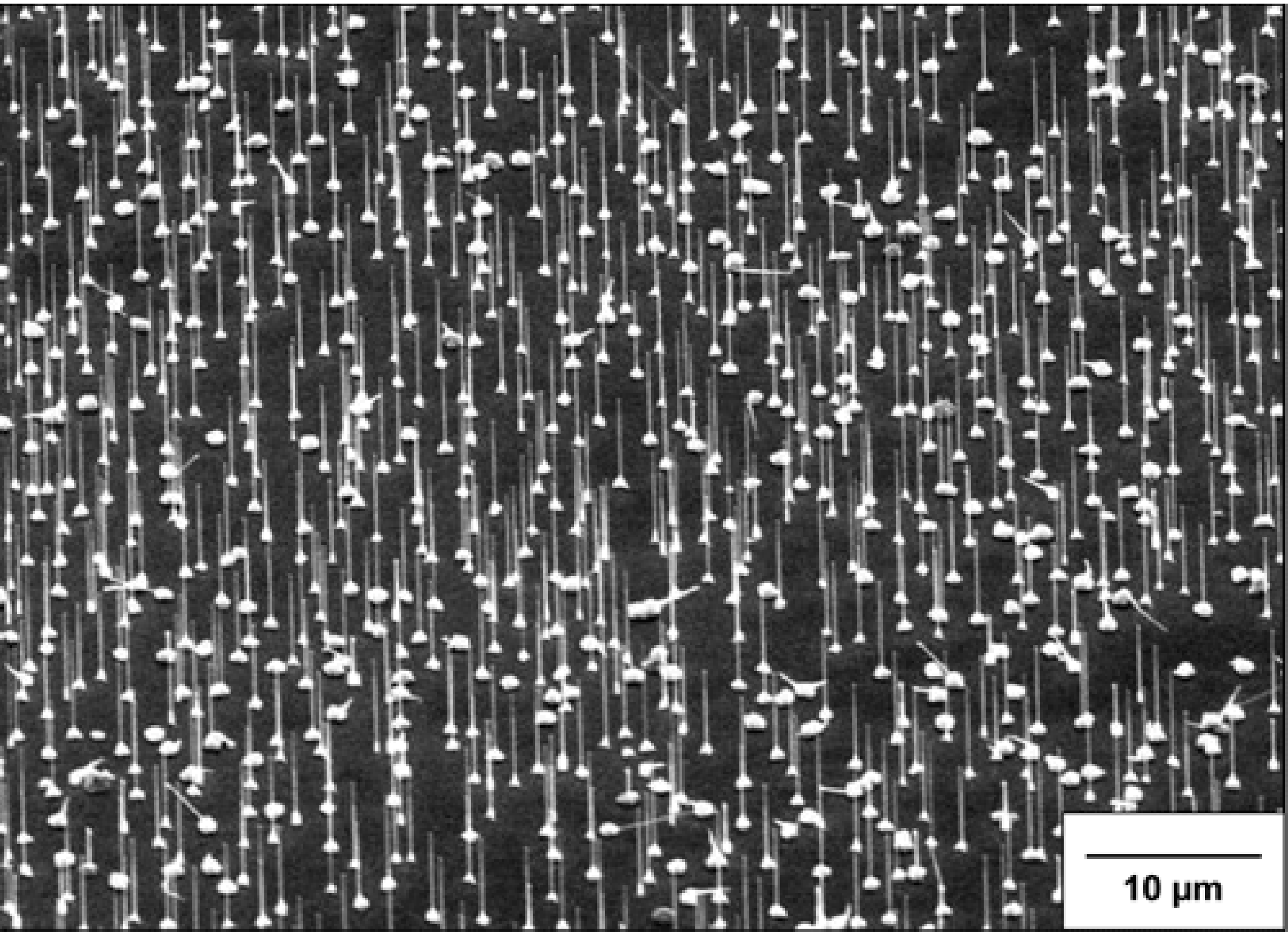}
\caption{Scanning electron microscope image of InAs nanowires grown in the cold-wall reactor. The image is taken at a substrate angle of 45$^\circ$. A ~5 mm $\times$ 5 mm InAs $<$111$>$B growth substrate is used.}
\end{figure}

A typical growth run with the cold-wall reactor consists of the following steps:
\begin{enumerate}
\item Sample preparation-- The sample preparation and application of the gold colloid is identical to the procedure detailed in the hot-wall growth section.
\item Sample transfer-- The sample is placed inside the load-lock, which is then evacuated to a pressure of $<$ 6 $\times$ 10$^{-6}$ Torr. Once this pressure is achieved, the gate valve that separates the load-lock from the growth chamber is opened. The quartz sample holder is then lowered onto the Meivac heater block, the transfer arm is extracted, and the gate valve is closed.
\item Growth preparation-- The precursor valves are opened and the system pressure stabilized as described for the hot-wall growth. The typical growth pressure is 50 Torr.
\item Sample growth-- The sample is first annealed at 600 $^\circ$C under a TEAs flow of 5 SCCM. We then use a two-temperature growth method to produce high quality nanowires.\cite{Joyce_NanoLett_2007} The temperature is lowered to 500 $^\circ$C, and a TMI flow of 1.25 SCCM is turned on for a brief 5 min nucleation step. The temperature is then lowered to 440 $^\circ$C for an extended growth step. For each step the precursor partial pressures are p$_{TMI}$ = 4.6 $\times$ 10$^{-6}$ atm and p$_{TEAs}$ = 5.4 $\times$ 10$^{-5}$ atm. After the growth is complete, the TMI flow is turned off, and the sample is cooled to 150 $^\circ$C. Finally, the flow of hydrogen is turned off.
\item Sample transfer-- The chamber is roughed to a pressure of 350 mTorr with the mechanical pump. Next, the chamber is turbo pumped to $<$ 1 $\times$ 10$^{-6}$ Torr, and the load-lock pumped to $<$ 6 $\times$ 10$^{-6}$ Torr. When the pressures are low enough, the gate valve is opened, and the sample is removed by lifting the quartz holder off the heater with the transfer fork. After the sample is safely in the load-lock, the gate valve is closed, and the sample is removed for inspection.
\item Placement into standby mode-- The growth chamber is left with its turbo pump on and eventually achieves its base pressure. The load lock is left vented.
\end{enumerate}

\begin{figure}[b]
%\centering
\includegraphics[width=8.6cm]{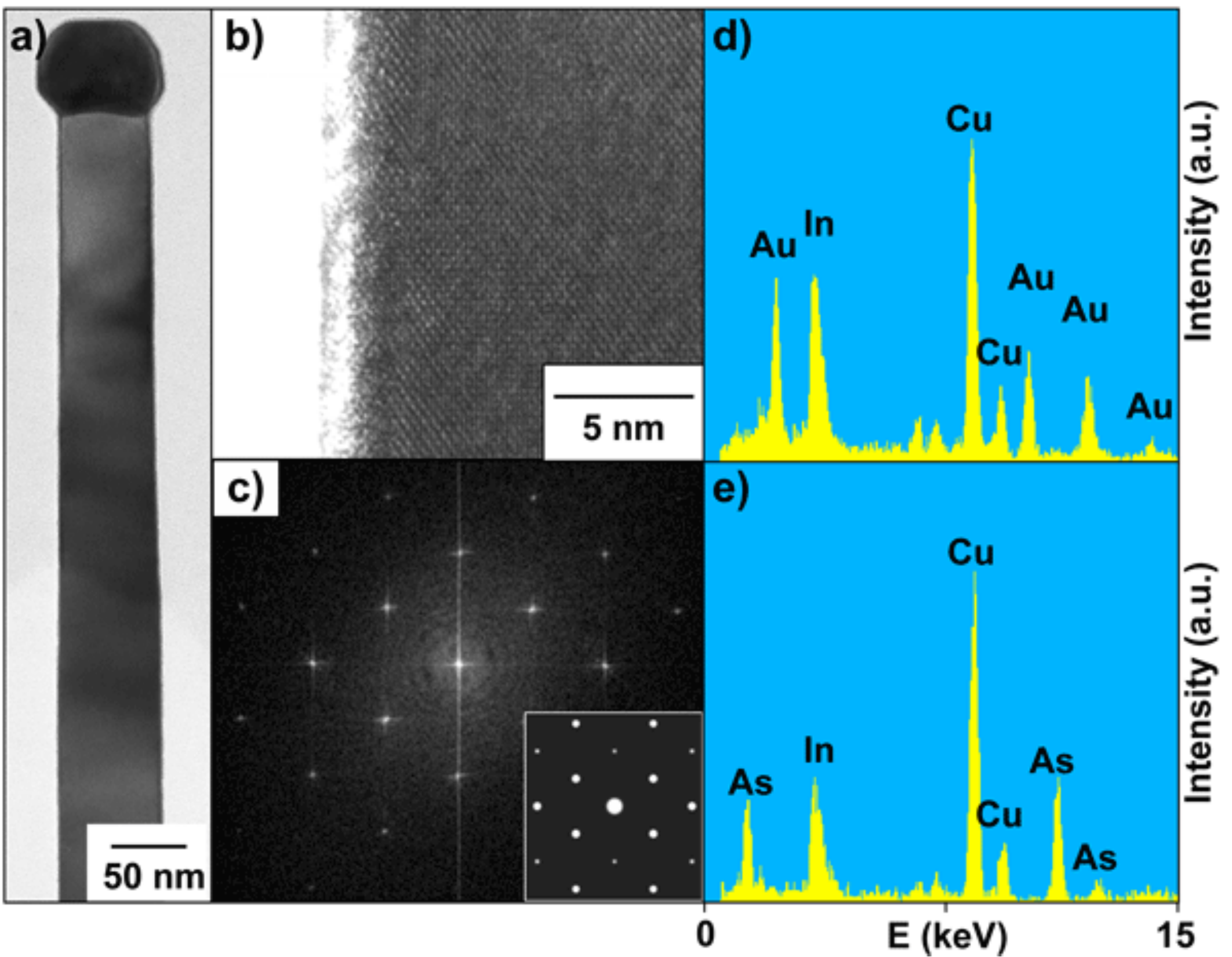}
\caption{High resolution microscopy of InAs nanowires grown in the cold-wall reactor. (a) TEM image of an InAs nanowire. The Au catalyst particle is visible on the end of the wire. A small amount of tapering ($\sim$15 nm in diameter per micron in length) is due to sidewall growth. (b) A $\sim$1 nm thick native oxide layer covers the wire and is visible in a HRTEM image. The wire is largely free of defects. (c) A Fourier transform of the HRTEM image is consistent with the diffraction pattern expected from an InAs wire with zincblende structure. Inset: Predicted diffraction pattern calculated using CrystalMaker software.\cite{crystalmaker} (d) Energy dispersive x-ray analysis of Au catalyst particle. (e) Energy dispersive x-ray analysis of InAs nanowire. Copper peaks are due to the TEM grid supporting the wire.}
\end{figure}

Growth runs with samples at the center and the edges of the heater confirm that the temperature and flow uniformity of our reactor doesn't equal that of commercial products, as the growth rate and morphology vary significantly as a function of position.  However, growths with small samples positioned at the center of the heater were found to be very reproducible, with slight run-to-run variations in growth rate but not morphology. Figure 5 shows a scanning electron microscope (SEM) image of InAs nanowires grown at the center of the heater. Nearly all of the wires grow in a vertical orientation, with a high areal density. Table 1 compares the uniformity of nanowires grown in this reactor to the hot-wall reactor.  High resolution electron microscopy was used to study the physical properties of individual nanowires. A bright field transmission electron microscope (TEM) image of a single InAs nanowire is shown in Fig.\ 6(a). The gold catalyst particle is clearly visible in this image at the top of the wire. In addition, we measure the degree of tapering (diameter change per micrometer of nanowire length) to be ~15 nm/$\mu$m. The high resolution TEM (HRTEM) image of the nanowire in Fig.\ 6(b) demonstrates high crystal quality with very few twining defects. A Fourier transform of the HRTEM image is consistent with the diffraction pattern expected from an InAs wire with zincblende structure [see Fig.\ 6(c)]. Energy dispersive x-ray analysis of the gold catalyst particle, shown in Fig.\ 6(d), indicates that it contains In, consistent with the VLS growth mechanism. Spectra shown in Fig.\ 6(e) are taken from a segment of wire and confirm that the chemical composition is indeed InAs. Copper peaks in both spectra are due to the TEM grid supporting the wire.

\begin{figure}[b]
%\centering
\includegraphics[width=8.6cm]{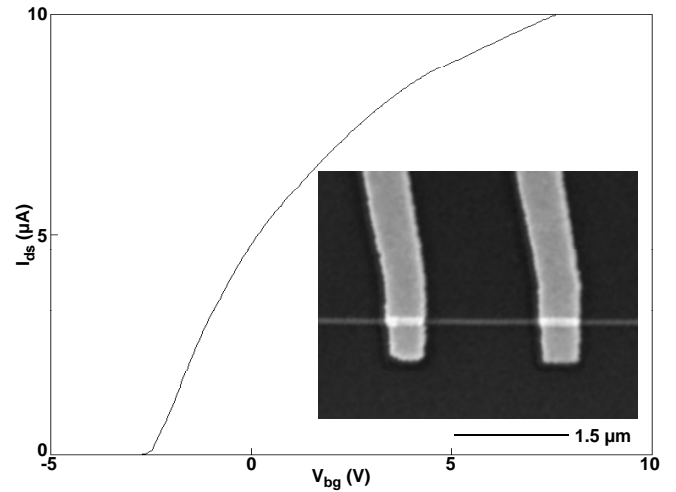}
\caption{Current, I$_{ds}$, through an InAs nanowire as a function of back gate voltage, V$_{bg}$. Measurements were taken at a temperature of 4.2 K with a source-drain bias of 100 mV. The wire shows clear field effect gating and is n-type. Inset: scanning electron microscope image of a back-gated InAs nanowire field effect transistor. Titanium/Gold source and drain contacts are fabricated using electron beam lithography. The wire was deposited on a thermally oxidized Si substrate.}
\end{figure}

Transport properties of the InAs nanowires were investigated by fabricating simple backgated FETs, as illustrated in Fig.\ 7. InAs nanowires were dispersed in ethanol and then dropcast onto an oxidized Si substrate. Source and drain contacts were defined using electron beam lithography. To lower the contact resistance, the native oxide on the surface of the InAs nanowire is removed using an ammonium polysulfide etch immediately before evaporation of the Ti/Au ohmic contacts.\cite{Suyatin_Nanotech_2007} Samples are then annealed at 200 $^\circ$C for 1 minute. Measurements from a typical device are shown in Fig.\ 7. Current through the wire, I$_{ds}$, is measured as a function of backgate voltage,  V$_{bg}$, for a source-drain bias V$_{ds}$=100 mV. The nominally undoped wire exhibits n-type behavior, consistent with results obtained from InAs nanowires grown in commercial MOVPE reactors.\cite{Wei_NanoLett_2009} From the transconductance, we extract a 4.2 K field effect mobility of $\sim$16,000 cm$^2$/(V$\cdot$s) for this wire. Measurements from 37 samples yield mobilities in the range of 4000--18000 cm$^2$/(V s), consistent with values reported in the literature.

\section{VI. CONCLUSIONS}
We have presented detailed descriptions of the assembly and operation of two different types of MOVPE nanowire growth systems. An inexpensive hot-wall reactor yielded InAs nanowire growth for a narrow range of operating conditions, but construction of a cold-wall reactor increased the range of parameter space over which we were able to obtain reproducible nanowire growth. InAs wires, microns long, and $\sim$60 nm in diameter, were grown in this cold-wall system. FETs with mobilities of 16000 cm$^2$/(Vs) were fabricated using the wires, demonstrating high quality electronic properties. Other material systems can be explored by simply changing the metal organic precursors used in the gas handling system.

\section{ACKNOWLEDGEMENTS}
We thank the Ensslin Group at ETH Zurich, and Stephen Weinman and Chun Yu of the Hongkun Park Group at Harvard, for invaluable advice on the design and operation of the hot-wall reactor. Prof.\ Jeff Schwartz's assistance with pyrophoric chemical transfers is appreciated. We also thank Jim Boehlert, Geoff Gettelfinger, and Prof.\ Jim Sturm for advice on building an exhaust scrubber system. This research was supported by the Sloan Foundation, the Packard Foundation, the Army Research Office, and the NSF funded Princeton Center for Complex Materials (Grant DMR-0819860).

\end{document}